\documentclass[aps,prl,twocolumn,showpacs,showkeys,floatfix]{revtex4}

\usepackage{graphicx}
\usepackage{dcolumn}
\usepackage{bm}

\newcommand{\Ei}{\mbox{Ei}}

\newcommand{\diffl}[2]{\frac{d #1}{d #2}}

\begin{document}


\title{Approximate, analytic solutions of the Bethe equation for
   charged particle range}

\date{December 16, 2008, revised February 10, 2009 -- LLNL-JRNL-410093}

\author{Damian C. Swift}
\email{dswift@llnl.gov}
\affiliation{%
   PLS-CMMD, Lawrence Livermore National Laboratory,
   7000 East Avenue, Livermore, California 94550, USA
}

\author{James M. McNaney}
\affiliation{%
   PLS-CMMD, Lawrence Livermore National Laboratory,
   7000 East Avenue, Livermore, California 94550, USA
}

\begin{abstract}
By either performing a Taylor expansion or making a polynomial approximation,
the Bethe equation for charged particle stopping power in matter
can be integrated analytically to obtain the range of charged particles
in the continuous deceleration approximation.
Ranges match reference data to the expected accuracy of the Bethe model.
In the non-relativistic limit, the energy deposition rate was also found
analytically.
The analytic relations can be used to complement and validate numerical
solutions including more detailed physics.
\end{abstract}


\maketitle

\section{Introduction}
The deceleration of ions as they pass through matter is important in a wide
range of fields:
medical ion radiation therapy, such as the treatment of tumors
\cite{Brahme2004};
radiography with ions \cite{Li2006};
radiolysis of chemical compounds \cite{Chitose1999};
ion implantation in material processing and semiconductor doping 
\cite{Shockley1954};
gas discharge plasmas \cite{Tsendin1995};
locality of energy deposition in nuclear fusion plasmas, including
ion beam heating for controlled thermonuclear fusion \cite{Keefe1982};
the design of radiation shielding for nuclear reactors \cite{Normand1989} and 
spacecraft \cite{Wilson1997}; 
and particle physics experiments \cite{Mulhearn2004}.
Energy loss for different combinations of ion specie, ion energy,
and decelerating material has been measured since the early 20th century 
\cite{Crowther1906}, with a corresponding development in theoretical work.
Calculations of ion energy loss, and hence range, are very frequently 
performed using variants of the Bethe \cite{Bethe1930} and
Bethe-Bloch \cite{Bloch1933} relations.
Experimental and theoretical developments have focused
on making corrections to the original Bethe relation to account for details of
interactions with bound electrons and crystal structures at low energies
\cite{Barkas1956,Ziegler1985},
and quantum-mechanical limits to the transfer of energy under extreme
relativistic conditions \cite{Jackson1999}.
The original Bethe relation is still used widely, particularly when 
a reliable, approximate result is needed rapidly, or in the fairly wide
range of energies where the Bethe relation is adequately accurate
\cite{Bichsel2004}.

The Bethe relation describes the stopping power:
the rate at which a moving ion loses energy to the surrounding material.
It is not trivial to use this relation to obtain the range
of an ion, i.e. the distance for it to lose all of its kinetic energy.
In practice, ion ranges are calculated using numerical integration
in multi-physics computer programs,
or from scaling laws normalized to the range for other
energies or masses.
However, reliance on sophisticated computer programs for infrequent
calculations, without the ability to make a compact analytic estimate, 
can lead to errors.
Analytic solutions are also valuable for validating computer programs
reproducing the same physics.
Here we point out analytic solutions to accurate approximations of the
Bethe relation, which can give good estimates of ion ranges in matter.

\section{Range from stopping power}
In the continuous deceleration approximation, charged particles
traversing matter lose kinetic energy $E$ at a rate depending 
on their instantaneous energy and the local material.
Expressed as the energy loss rate per distance traveled, the stopping
power $dE/dx$ can be used to determine the range $l$ of the particle,
by integrating the deposition until the particle is stationary:
\begin{equation}
\int_0^l\diffl{E(x)}x\,dx = -E_0.
\end{equation}
However, $dE/dx$ is expressed naturally in terms of $E$ rather than $x$.
Rearranging,
\begin{equation}
l=\int_{E_0}^{0}\diffl xE\,dE.
\label{eq:range}
\end{equation}
The integral can be found numerically for arbitrary stopping powers, 
or analytically for stopping powers of sufficiently simple form.

The Bethe equation \cite{Bethe1930} describes the deceleration of
charged particles by interaction with the electrons in matter:
\begin{equation}
\diffl Ex = 
-\frac{4\pi}{m_e c^2} \frac{N Z z^2}{\beta^2}
\left(\frac{q^2}{4\pi\epsilon_0}\right)^2
\left[
\ln\frac{2 m_e c^2\beta^2}{\bar I\left(1-\beta^2\right)}-\beta^2
\right]
\label{eq:bethe}
\end{equation}
where $\beta = v/c$, $v$ is the ion speed,
$\bar I$ is the effective ionization of the target material,
$Z$ and $z$ are the atomic numbers of the target and ion species respectively,
$N$ is the number density of target nuclei,
$m_e$ is the mass of an electron,
$\epsilon_0$ is the permittivity of free space, and $c$ is the speed of light.

To find the range of charged particles from Eq.~\ref{eq:bethe}, 
Eq.~\ref{eq:range} can be integrated numerically, though this is not
straightforward because of a singularity at low energies.
We have not found an analytic solution for the integral.

\section{Taylor expansion}
$-dx/dE$ can be expanded as a Taylor series to make it more tractable
for integration.
This can be done for Eq.~\ref{eq:bethe} with a relativistic expression
for $\beta(E)$;
we do it also for a non-relativistic $\beta(E)$ because the
resulting integral is more amenable to subsequent manipulation.

\subsection{Relativistic}
The relativistic relation between $\beta$ and kinetic energy $E$ is
\begin{equation}
\beta(E) = \frac{\sqrt{E(E+2m_ic^2)}}{E+m_i c^2}
\label{eq:betarel}
\end{equation}
where $m_i$ is the rest-mass of the moving ion.
For later convenience, we scale key quantities to be dimensionless.
Substituting into Eq.~\ref{eq:bethe} and expanding about zero,
\begin{equation}
-\diffl xE=
\frac{4\pi\epsilon_0^2 m_e c^2}{q^4 z^2 N Z}\left[
\frac{2\hat E}L-
\frac{3\hat E^2\left(L-1\right)}{L^2}
\right]+O(\hat E^3)
\label{eq:relseries}
\end{equation}
where
\begin{equation}
\hat E\equiv \frac E{m_i c^2}, \quad 
\hat I\equiv \frac {\bar I}{2 m_e c^2}
\end{equation}
are the scaled kinetic energy and mean ionization, and
\begin{equation}
L\equiv\ln\frac{2\hat E}{\hat I}.
\end{equation}
Integrating Eq.~\ref{eq:range}, the range is
\begin{equation}
l=\frac{4\pi\epsilon_0^2 m_e c^2 m_i c^2}{q^4 z^2 N Z}\left[
\frac 12\hat I^2\Ei\left(2L\right)
+\frac 34\hat I^3\Ei\left(3L\right)
-\frac{3 \hat E^3}L
\right]
\label{eq:relrange}
\end{equation}
where $\Ei(z)$ is the exponential integral function,
\begin{equation}
\Ei(z)\equiv-\int_{-z}^\infty\frac{e^{-t}}t\,dt.
\end{equation}

\subsection{Non-relativistic}
In the non-relativistic limit,
$\beta(E) = \sqrt{2 E/m_i}$.
Following the same procedure as above,
we find that the non-relativistic form of each of
$-dx/dE$ and $l$
is simply the first term of the corresponding relativistic relation.

\section{Mixed-species targets}
For a target comprising multiple elements, the stopping power can be
estimated from the combination of stopping powers from each element $s$
\begin{equation}
\diffl Ex \simeq \sum_s \diffl Ex(Z_s,N_s)
\label{eq:bragg_addition}
\end{equation}
-- the Bragg addition rule.
This relation is approximate because of chemical bond formation,
which alters the effective ionization.
For ion energies much greater than the bond energies, the approximation
should be accurate.

Expanding as before and gathering terms, the range can be expressed very
similarly to that for single-element targets.
In the relativistic case, 
\begin{equation}
l=\frac{4\pi\epsilon_0^2 m_e c^2 m_i c^2}{q^4 z^2 \tilde Z}\left[
\frac 12\tilde I^2\Ei(2\tilde L)
+\frac 34\tilde I^3\Ei(3\tilde L)
-\frac{3 \hat E^3}{\tilde L}
\right]
\end{equation}
where
\begin{equation}
\tilde I \equiv \exp\frac{\sum_s N_s Z_s \ln\hat I_s}{\tilde Z}, \quad
\tilde L \equiv \ln\frac{2 \hat E}{\tilde I} \\
\end{equation}
and
\begin{equation}
\tilde Z \equiv \sum_s N_s Z_s
\end{equation}
is the total electron density in the target.

In the non-relativistic case, 
\begin{equation}
l=\frac{2\pi\epsilon_0^2 m_e c^2 m_i c^2}{q^4 z^2 \tilde Z}
\tilde I^2\Ei\left(2\tilde L\right),
\end{equation}
which is again simply the first term of the relativistic relation.

\section{Polynomial fit to the stopping distance scale}
\begin{figure}
\begin{center}\includegraphics[scale=0.72]{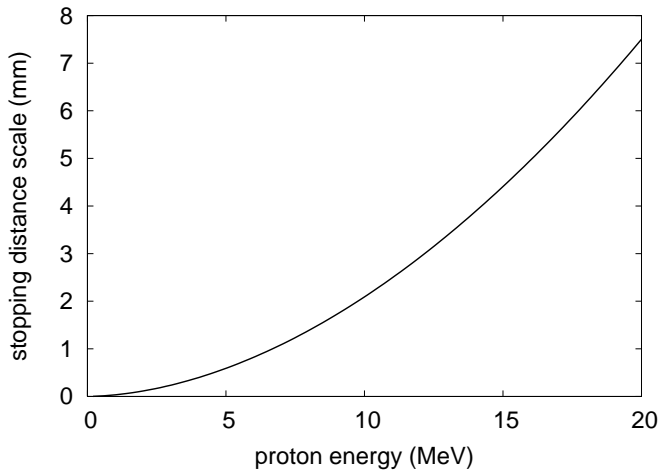}\end{center}
\caption{Example calculation of stopping distance scale:
   protons in water. The curve can be fitted well by a quadratic.}
\label{fig:stopscale}
\end{figure}

Although well-characterized numerical approximations to 
the exponential integral exist \cite{Pecina1986},
they are not available as standard functions in mainstream computer 
languages, and require significant effort to implement from scratch.
However, the logarithmic terms in the stopping power and range vary slowly
compared with the powers of $E$;
over wide ranges of energy, the stopping power can be approximated
accurately by low-order polynomials.
The use of polynomial approximations avoids the need to evaluate the
exponential integral function.
We define a stopping distance scale
\begin{equation}
D\equiv-Edx/dE,
\end{equation}
which is particularly well-behaved above the low-energy singularity 
as it tends to zero with $E$, and increases monotonically
(Fig.~\ref{fig:stopscale}).
This quantity can be used as a crude, $O(1)$, (over)estimate of particle range,
without requiring any series expansion or integration.
Approximating $D$ by a polynomial
\begin{equation}
D_p(E) = \sum_j a_j E^j,
\end{equation}
the range (Eq.~\ref{eq:range}) is simply
\begin{equation}
l_p \simeq a_0\ln E+\sum_{j>0} \frac{a_j E^j}j.
\label{eq:polyrange}
\end{equation}
$a_0$ must be zero, since $E^n/\ln(\alpha E)\rightarrow 0$ as $E\rightarrow 0$.

To find the charged particle range in a specific substance,
it is straightforward to tabulate $D(E)$ using Eq.~\ref{eq:bethe}
(and Eq.~\ref{eq:bragg_addition} for a multi-species target),
fit a polynomial $D_p(E)$, and evaluate Eq.~\ref{eq:polyrange}.
However, it is also possible to find a universal polynomial fit.
Defining for convenience a different scaled energy
\begin{equation}
F\equiv\frac{4 m_e E}{\bar I m_i} = \frac{2\hat E}{\hat I}
\end{equation}
the stopping distance scale is (using Eq.~\ref{eq:relseries})
\begin{equation}
D(F)=\frac{\pi \epsilon_0^2}{2 m_e q^4}\frac{m_i \bar I^2}{z^2 N Z}
\left[
\frac{F^2}{\ln F}+\frac 38\frac{\bar I}{m_e c^2}\frac{1-\ln F}{(\ln F)^2}F^3
\right],
\end{equation}
where the first term is the non-relativistic approximation.
The prefactor comprises universal constants and a simple problem-specific factor
$m_i \bar I^2/z^2 N Z$.
The relative magnitude of the relativistic term to the non-relativistic term
depends only on $\bar I$.
Thus, by finding polynomial approximations
\begin{eqnarray}
-\frac{F^2}{\ln F} \simeq P_{NR}(F)=\sum_j n_j F^j \\
\frac{\ln F-1}{(\ln F)^2} F^3 \simeq P_R(F)=\sum_j r_j F^j
\end{eqnarray}
it is straightforward to find the polynomial coefficients for any problem:
\begin{equation}
a_j = \frac{\pi \epsilon_0^2}{2 m_e q^4}\frac{m_i \bar I^2}{z^2 N Z}
\left(\frac{4 m_e}{\bar I m_i}\right)^j
\left(
n_j
+\frac 38\frac{\bar I}{m_e c^2}r_j
\right)
\end{equation}
and hence the range through Eq.~\ref{eq:polyrange}.

The derivation presented above is valid for any choice of units.
The Bethe relation breaks down when the logarithm changes sign,
i.e. when $E$ approaches $\bar I m_i/4 m_e$.
Physically meaningful distances are obtained for {\it greater} energies.
Using the
Bloch estimate \cite{Bloch1933} for the effective ionization of the material,
\begin{equation}
\bar I \simeq 10 Z q,
\label{eq:bloch}
\end{equation}
the relations are valid for $E > 4600 Z q$.
Polynomial fits were calculated over a range suitable for hadrons of energy 
$\sim$MeV to GeV (Table~\ref{tab:polyfits}).
The relativistic term diverges rapidly outside the fitting region.

\begin{table}
\caption{Polynomial fits to stopping distance scale functions.}
\label{tab:polyfits}
\begin{center}
\begin{tabular}{|c|l|l|}\hline
{\bf parameter} & $5\le F\le 100$ & $100\le F\le 10000$ \\ \hline
$n_1$ & $ 2.17423$ & $ 1.64377\times 10$ \\
$n_2$ & $ 2.29035\times 10^{-1}$ & $ 1.31696\times 10^{-1}$ \\
$n_3$ & $-3.36317\times 10^{-4}$ & $-4.55336\times 10^{-6}$ \\
$n_4$ & & $ 2.07676\times 10^{-10}$ \\
\hline
\hline
 & $10\le F\le 200$ & $500\le F\le 10000$ \\ \hline
$r_2$ & $-1.33946\times 10$ & $-9.92931\times 10^3$ \\
$r_4$ & $ 2.04195$ & $ 3.49521\times 10$ \\
$r_6$ & $ 1.58588\times 10^{-1}$ & $ 1.00674\times 10^{-1}$ \\
$r_8$ & $-7.67337\times 10^{-5}$ & $-7.28447\times 10^{-7}$ \\
\hline\end{tabular}\end{center}\end{table}

\section{Energy deposition profile}
Given $E(x)$, the profile of energy deposition $-dE(x)/dx$ (Bragg curve)
can be calculated.
$E(x)$ is the inverse of the range, $l(E)$.

\begin{figure}
\begin{center}\includegraphics[scale=0.72]{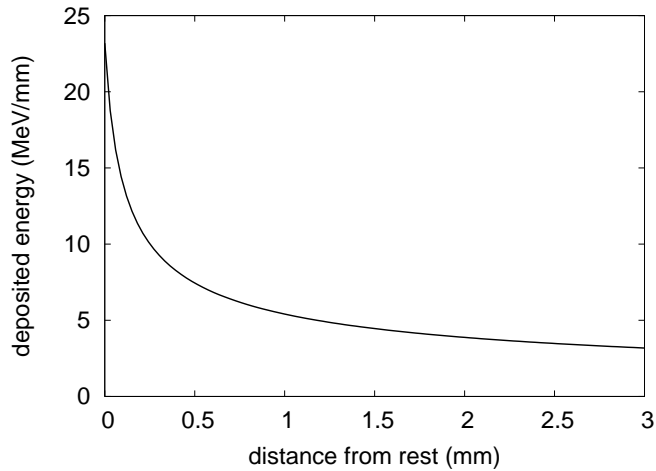}\end{center}
\caption{Example calculation of Bragg curve via a polynomial
   (quadratic) fit to $D(E)$: protons in water.
   The curve is presented backward from its usual form:
   as if the particles are accelerating from rest.
   Conceptually, the curve can be continued to arbitrarily high energies,
   i.e. long ranges.}
\label{fig:bragg}
\end{figure}

For the non-relativistic range relation, 
$E(x)$ can be expressed in terms of the inverse of the exponential
integral function,
\begin{equation}
-\diffl{E(x)}x\simeq \frac{q^4z^2NZ}{4\pi\epsilon_0^2\bar I}
\exp\left[\frac 12\Ei^{-1}(\alpha x)\right]\Ei^{-1}(\alpha x)
\end{equation}
where
\begin{equation}
\alpha\equiv\frac{2 m_e q^4 z^2 NZ}{\pi\epsilon_0^2\bar I^2 m_i}.
\end{equation}
If the stopping distance scale is represented locally in energy by 
a sufficiently simple polynomial,
then it too may be used to calculate $-dE/dx$.
For example, taking a local quadratic fit
\begin{equation}
D_p = a_1 E+a_2 E^2
\quad\Rightarrow\quad
l_p = a_1 E+\frac{a_2}2E^2,
\end{equation}
one obtains
\begin{equation}
-\diffl{E}x\simeq \frac 1{\sqrt{a_1^2+2 a_2 x}}
\end{equation}
(Fig.~\ref{fig:bragg}).

\section{Example calculations}
Reference calculations are used to validate radiation protection
simulations using different computer programs.
Here we compare the analytic solutions of the Bethe relation with
results from widely-used programs SRIM \cite{SRIM},
which uses numerical solutions of more detailed stopping powers
developed from the Bethe relation, and MCNP \cite{MCNP},
which collects Monte-Carlo statistics for the simulated interaction of 
individual particles.
Trial calculations were made for protons and $\alpha$-particles stopping in
Al and water.

We use the Bloch estimate, Eq.~\ref{eq:bloch} for the effective ionization
of the target material.
More accurate calculations have been developed more recently, but the
original Bloch estimate serves to demonstrate the correctness of our
analysis.

The results are consistent with the accuracy of the Bethe relation
itself (Table~\ref{tab:comparison}),
and are consistent with direct numerical integration of the Bethe equation 
without being affected by the low energy singularity.
The greatest difference was for relativistic protons in Al.
In this regime, radiative losses and nuclear reactions
become significant \cite{Bichsel2004} and the Bethe relation requires
additional corrections.

\begin{table}
\caption{Comparison between analytic calculation and computer simulations
   of ion ranges (in millimeters).}
\label{tab:comparison}
\begin{center}
\begin{tabular}{|l|r|r|r|}\hline
{\bf system} & {\bf analytic} & {\bf MCNP 5} & {\bf SRIM} \\ \hline
\hline
p $\rightarrow$ Al & & & \\
10 MeV & 0.59 & 0.62 & 0.63 \\
100 MeV & 36 & 37 & 37 \\
1 GeV & 180 & 1510 & 1530 \\
\hline
$\alpha$ $\rightarrow$ Al & & & \\
10 MeV & 0.054 & 0.062 & 0.061 \\
100 MeV & 3.0 & 3.2 & 3.1 \\
1 GeV & 170 & 180 & 180 \\
\hline
p $\rightarrow$ water & & & \\
10 MeV & 1.1 & 1.2 & 1.2 \\
100 MeV & 73 & 77 & 76 \\
1 GeV & 300 & 320 & 320 \\
\hline
$\alpha$ $\rightarrow$ water & & & \\
10 MeV & 0.097 & 0.110 & 0.110 \\
100 MeV & 6.0 & 6.3 & 6.2 \\
1 GeV & 350 & 380 & 375 \\
\hline\end{tabular}\end{center}\end{table}

\section{Conclusions}
Analytic solutions were found to power series expansions and polynomial fits
to the Bethe relation.
These solutions provide a convenient way to calculate ion ranges and
energy deposition in regimes where the Bethe relation is valid,
i.e. kinetic energies of roughly 1-100\,MeV/u,
without depending on numerical integration.
The use of a Taylor series restricts the accuracy at high energy;
the relativistic expansion thus incorporates relativistic contributions to the
range but is not valid to arbitrarily high energies.
However, the analytic solutions can readily be used with more accurate 
formulations of the effective ionization.
The accuracy was demonstrated by comparison with simulations from widely-used
computer programs of ion ranges in Al and water.

\section*{Acknowledgments}
The authors would like to acknowledge the contribution of Tim Goorley
(Los Alamos National Laboratory) for providing reference calculations
of stopping distance using SRIM and MCNP,
and Sergei Kucheyev (Lawrence Livermore National Laboratory) for helpful
discussions.
This work was performed in support of Laboratory-Directed
Research and Development project 09-ERD-037 under the auspices
of the U.S. Department of Energy under contract DE-AC52-07NA27344.

\end{document}